\documentclass[twocolumn,10pt]{IEEEtran}
\usepackage{amsmath,amssymb,mathrsfs}
\usepackage{graphicx}
\usepackage{psfrag}
\usepackage{subfigure}
\usepackage{cite}
\usepackage{multirow}
\usepackage{stfloats}
\usepackage{color}
\usepackage{amsmath}
\usepackage[harvard,dcucite]{harvard}
\usepackage{soul}
\usepackage{color, soul}
\title{On the Estimation and Control of Human Body Composition}

\author{
Mahmood Karimi and Ramesh R. Rao\\
Qualcomm Institute, Calit2, University of California, San Diego\\
La Jolla, CA 92093, USA\\
Email: \{makarimi, rrao\}@ucsd.edu
}
\begin{document}

\maketitle    

%%%%%%%%%%%%%%%%%%%%%%%%%%%%%%%%%%%%%%%%%%%%%%%%%%%%%%%%%%%%%%%%%%%%%%
\begin{abstract}
Obesity is a chronic disease that can lead to an increased risk of other serious chronic diseases and even death. We present switching and time-delayed feedback-based model-free control methods for the dynamic management of body mass and its major components. The estimation of body composition based on human body weight dynamics is proposed using a soft switching-based observer. Additionally, this paper addresses the control allocation problem for optimal body weight management using linear algebraic equivalence of the nonlinear controllers based on dynamic behavior of body composition described in literature. A control allocator system computes the required energy intake and energy expenditure from a controlling range of inputs to track a desired trajectory of body mass by optimizing a weighted quadratic function. Simulation results validate the performance of the proposed controllers and the observer under disturbances in recording energy intake and energy expenditure figures.

\vspace{3mm}
\end{abstract}

%%%%%%%%%%%%%%%%%%%%%%%%%%%%%%%%%%%%%%%%%%%%%%%%%%%%%%%%%%%%%%%%%%%%%%
\section{INTRODUCTION}

Obesity, a serious social and public health problem, is increasing dramatically among the population not only in high-income and developed countries, but also in low-to-middle-income and poor, developing countries, particularly in urban settings. Obesity and being overweight result from a variety of factors, such as physical inactivity, high level of stress, as well as inappropriate diet. Obesity and being overweight are also associated with heart disease, certain types of cancer, type 2 diabetes, stroke, arthritis, breathing problems, and psychological disorders, such as depression \cite{Laila2010}, \cite{Sentocnik--AtanasijevicKunc--Drinovec--Pfeifer2013}.

Many studies have been conducted to try and understand the etiology of weight gain and obesity. In \cite{Chow--Hall2008}, a model based on macronutrient and energy flux balance is presented. A computational model, presented in \cite{Hall2010}, shows how diet perturbations result in adaptations of fuel selection and energy expenditure that predict body weight and composition changes in both obese and non-obese individuals. In \cite{NavarroBarrientos--Rivera--Collins2011}, the authors incorporated both physiological and psychological factors in a dynamical model to help develop behavioral interventions.

Solving the obesity problem through healthy body weight management has been of interest in the control community \cite{Karimi--Rao2015}. In \cite{Laila2010}, closed loop control of body composition was done, and the energy intake into the body is regarded as the input control to the system. Clinical open-loop and closed-loop control efforts for various scenarios were studied in \cite{Sentocnik--AtanasijevicKunc--Drinovec--Pfeifer2013}, and the efficacy of the treatment can, in this way, be significantly improved.

Control of body mass or its major components may need daily body composition measurement. Several methods, such as air displacement plethysmography, dual-energy x-ray absorptiometry, bioimpedance spectroscopy, quantitative magnetic resonance and magnetic resonance imaging/spectroscopy, are presented in \cite{Baracos2012}. These methods are expensive to administer and inconvenient to arrange for frequent measurements. To minimize the associated cost and inconvenience, we propose a technique for estimation of body composition using observer design based on daily measurements of total body weight alone.

Feedback linearization (FL) and sliding mode control (SMC) are two widely used control schemes for nonlinear systems. Despite their desirable features, the inability to handle explicit input constraints is a major drawback of using FL/SMC, while the body weight control system described in this research clearly has input constraint--limitations in energy intake and physical activity. To resolve this problem, \cite{Gwak--Masada2008} proposed a constrained FL/SMC nonlinear optimal controller that can be represented in a simple constrained linear least-square problem.

A precise dynamic model is difficult to build, and the identification of the model parameters has to be individualized and can vary over time. Therefore, a controller that does not require a dynamic model would be a useful and more robust tool than model-based controllers, especially when precise tracking is achievable with small control gains. 

In addition, measurements of energy intake and energy expenditure are prone to errors and lead to huge disturbances to the system \cite{Hall2015}. Hence, the robustness of an observer to disturbances is essential for accurate prediction of body composition.   

This paper, partly presented in \cite{Karimi--Rao2014}, is based on the dynamic behavior of body weight based on macronutrient intake and the energy balance model. Here, two model-free control algorithms for body weight management using \textit{a priori} knowledge of control input and a switching-based PI controller are presented. The energy expenditure of the body is regarded as the input control to the system. Body components are estimated using a dynamic observer based on daily measurement of total body weight supplemented by periodic measurement of individual body components. Next, control allocation problems are formulated using FL/SMC as linear algebraic equations, and input constraints and optimization issues are addressed. Finally, effectiveness and robustness of the proposed algorithms are demonstrated through simulations.

\section{PROBLEM DEFINITION}
The major components of human body mass are fat and lean masses and both of these factors as well as total body mass need to be managed.\\
\indent
The goal of this work is to propose controllers that suggest daily energy intake or physical activity to track a preferred trajectory of weight/fat reduction or increase. It is suitable for professional athletes, the elderly individuals at risk of Sarcopenia, people with disabilities or sickness and those who need to precisely manage their weight/fat/lean due to medical reasons or less body weight fluctuation.\\
\indent
Since the controllers are robust enough, some deviation from their suggestions is tolerable, and good tracking can be still practically achieved. So, the human preference issue is considered implicitly. \\
\indent
When it is intended to control fat/lean mass, we need to have daily estimation of fat/lean mass to use in the controller whether the controller is model-free or model-based. Since measurement of fat/lean mass is hard and expensive to do every day, we design an observer to estimate the body compositions using daily total body weight measurement and periodic fat mass measurement. The dynamic model, which encompasses fat and lean masses as system states, is used for observer design. 

It is assumed that the subject is able to follow the control suggestions and their moods, energy intake and physiological behavior are normal and not pathological. Also, it is assumed that energy intake and expenditure are reasonably precisely measurable.

\vspace{1mm}
\section{BODY WEIGHT DYNAMICS BASED ON THE ENERGY BALANCE MODEL} \label{sec:dynamics}

To describe the energy balance in humans, different mathematical models have been developed. A comprehensive three-compartment model is presented in \cite{Sentocnik--AtanasijevicKunc--Drinovec--Pfeifer2013} and \cite{NavarroBarrientos--Rivera--Collins2011}.

The daily energy-balance (EB) equation is described as
\begin{equation} \label{eq:EB}
EB=EI-EE
\end{equation}
where $EI$ represents the daily energy intake, $EE$ represents the daily energy expenditure, and $BM$ is the body mass of the person expressed in terms of fat (F), lean (L) and extracellular fluid (ECF).
\begin{equation} \label{eq:BM}
BM=F+L+ECF
\end{equation}

The energy intake is based on the consumed food and its caloric value
\begin{equation} \label{eq:EI}
EI=k_1ci+k_2fi+k_3pi
\end{equation}
where $ci$ indicates carbohydrate intake, $fi$ and $pi$ indicate the fat intake and the protein intake (all representing model inputs), respectively, and $k_i$s are constant coefficients.

To track fat mass, lean mass and extracellular fluid, the following balance equations were used:
\begin{equation} \label{eq:Fat}
\frac{dF}{dt}=\frac{1-r}{\rho_{F}}EB
\end{equation}
\begin{equation} \label{eq:Lean}
\frac{dL}{dt}=\frac{r}{\rho_{L}}EB
\end{equation}
\begin{equation} \label{eq:ECF}
\frac{dECF}{dt} = \frac{\rho_{w}}{Na} \left[ 
\zeta_{Na}(ECF_{init}-ECF) - \zeta_{ci} \left( 1- \frac{ci}{ci_b} \right)
\right]
\end{equation}
where $\rho_F$ and $\rho_L$ are the energy density of the body fat and lean muscle mass respectively, $\rho_{w}$ is the density of water, $ci_b$ is the baseline carbohydrate intake, and the ratio $r$ is the parameter describing the imbalance denoted by $EB$ to the compartments fat mass and lean mass. This parameter, $r$, was defined in \cite{Hall2007} based on the Forbes formula \cite{Forbes1987}, which was obtained after analyzing body composition data collected from many adults:
\begin{equation} \label{eq:r}
r=\frac{c}{c+F}; \quad c=kk \frac{\rho_{L}}{\rho_{F}}
\end{equation}
Forbes found that this relationship was similar whether weight loss was due to diet or exercise. Prolonged exercise or a significant change in the protein intake may cause a different relationship for $r$.

The daily energy expenditure is calculated as follows:
\begin{equation} \label{eq:EEgeneral}
EE = [PA] + [TEF] + [RMR]
\end{equation}
where $PA$ (input to the model) represents the energy spent on physical activity, $TEF$ is the thermic effect of food, and $RMR$ is the resting metabolic rate needed for basic physiological processes. Equation \eqref{eq:EEgeneral} can be also written as \vspace{-3mm}

\begin{equation} \label{eq:EE}
\small EE = [\delta \cdot BM] + [\beta \cdot EI] + \left[ 
K+\gamma_L L+\eta_L \dot{L}+ \gamma_F F+\eta_F \dot{F} 
\right] 
\end{equation} 
where $\delta = \frac{PA}{BM}$ represents the physical activity coefficient and the constant $K$ depends on the initial conditions and is calculated using Equation \eqref{eq:EB} and Eqs. \eqref{eq:Fat} -- \eqref{eq:ECF} in the steady state: $EB = \dot{F}=$ $\dot{L}=\dot{ECF}=0$

Therefore, for the steady-state conditions, Eq. \eqref{eq:EE} can be rewritten as follows:
\begin{equation} \label{eq:K}
K=(1-\beta) \bar{EI} - \gamma_L \bar{L} - \gamma_F \bar{F} -\bar{PA}
\end{equation}
where the bars over the variables indicate the steady state of the corresponding variables.

The obtained model is used for an optimal controller, observer design, and simulations, which will be presented in the following sections.

\section{MODEL-FREE CONTROL}

\vspace{1mm}
\section*{Controller Design} \label{sec:controller}

Let the control objective be steering any combination of body composition masses, including total body mass, to a desired reference value/trajectory without using the dynamic model in the controller. It is considered that physical activity is the only control input to the system and that it conforms to a desired shape. The body weight dynamics~\eqref{eq:Fat}-\eqref{eq:ECF} can be formed as
\begin{equation} \label{eq:fgu}
\dot{x} = h(x,y,u) = f(x,y)+g(x,y)u
\end{equation}
\begin{equation} \label{eq:y}
y=Cx
\end{equation}

in which $x=[F,~L,~ECF]$, $u=\delta$, $y$ is a controlled output, $C$ is a $1 \times 3$ matrix and
\begin{equation}
\small f(t,x,y)=
\begin{bmatrix}
(1-a-bF)(EI - [\beta \cdot EI] - \left[ K+\gamma_L L+ \gamma_F F \right]) \\
(a+b F)(EI - [\beta \cdot EI] - \left[ K+\gamma_L L+ \gamma_F F \right] ) \\
\frac{\rho_{w}}{Na} \left[ 
\zeta_{Na}(ECF_{init}-ECF) - \zeta_{ci} \left( 1- \frac{ci}{ci_b} \right)
\right]
\end{bmatrix}
\end{equation}

\begin{equation}
g(x,y)=
\begin{bmatrix}
-(1-a-b F)BM \\
-(a+b F)BM \\
0
\end{bmatrix}
\end{equation}

Differentiating the output \eqref{eq:y} with respect to $t$ yields
\begin{equation} \label{eq:ydot}
\dot{y} = Cf(x,y)+Cg(x,y)u
\end{equation}

Introducing from \eqref{eq:ydot} a constant, $\bar{g}$, representing the nominal value of $g$, we can rewrite \eqref{eq:ydot} into the following equation:
\begin{align}
\dot{y} &= \bar{g}u + [Cf + (Cg-\bar{g} ) u] \nonumber \\
&= \bar{g}u + H(t)
\end{align}

where $H(t)$ denotes the total nonlinearity, which is expressed as
\begin{equation} 
H(t) = Cf + (Cg-\bar{g} ) u
\end{equation}

The desired error dynamics is defined with an asymptotically stable linear time invariant system as in the following:
\begin{equation}
\dot{e_y} = \Pi e_y
\end{equation}

\begin{figure}
\begin{center}
\subfigure[]{
\small
\includegraphics[trim = 0in 0in 0in 0in, clip, scale=0.45]{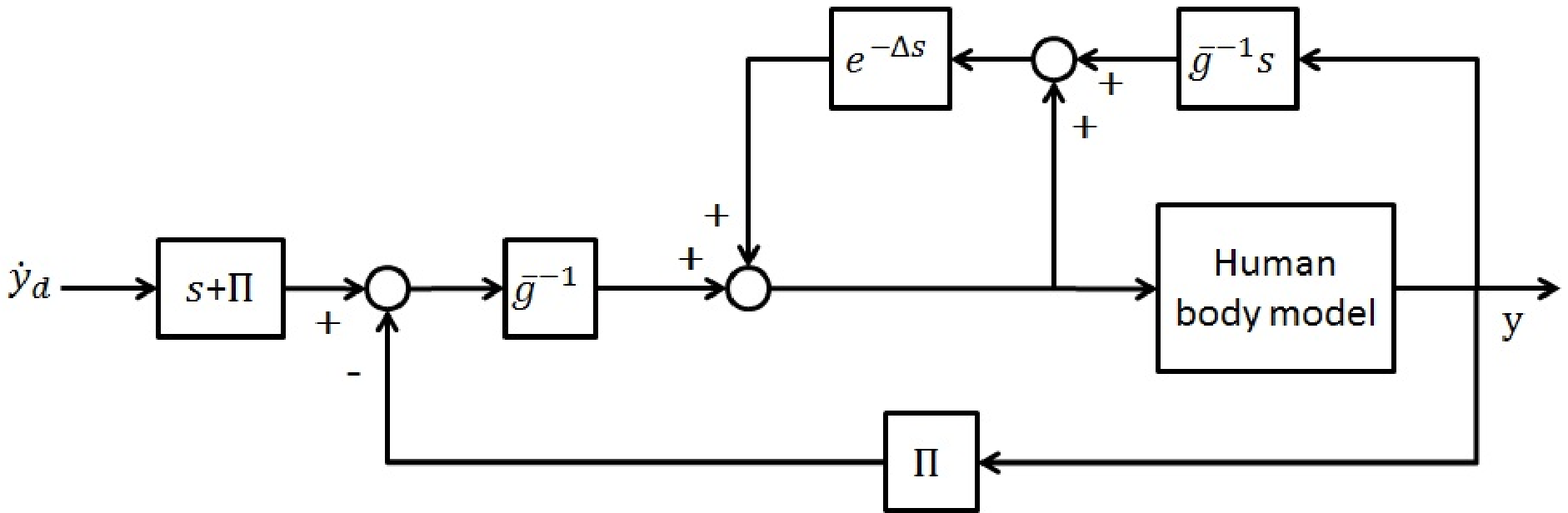}
}
\subfigure[]{
\small
\includegraphics[trim = 0in 0in 0in 0in, clip, scale=0.5]{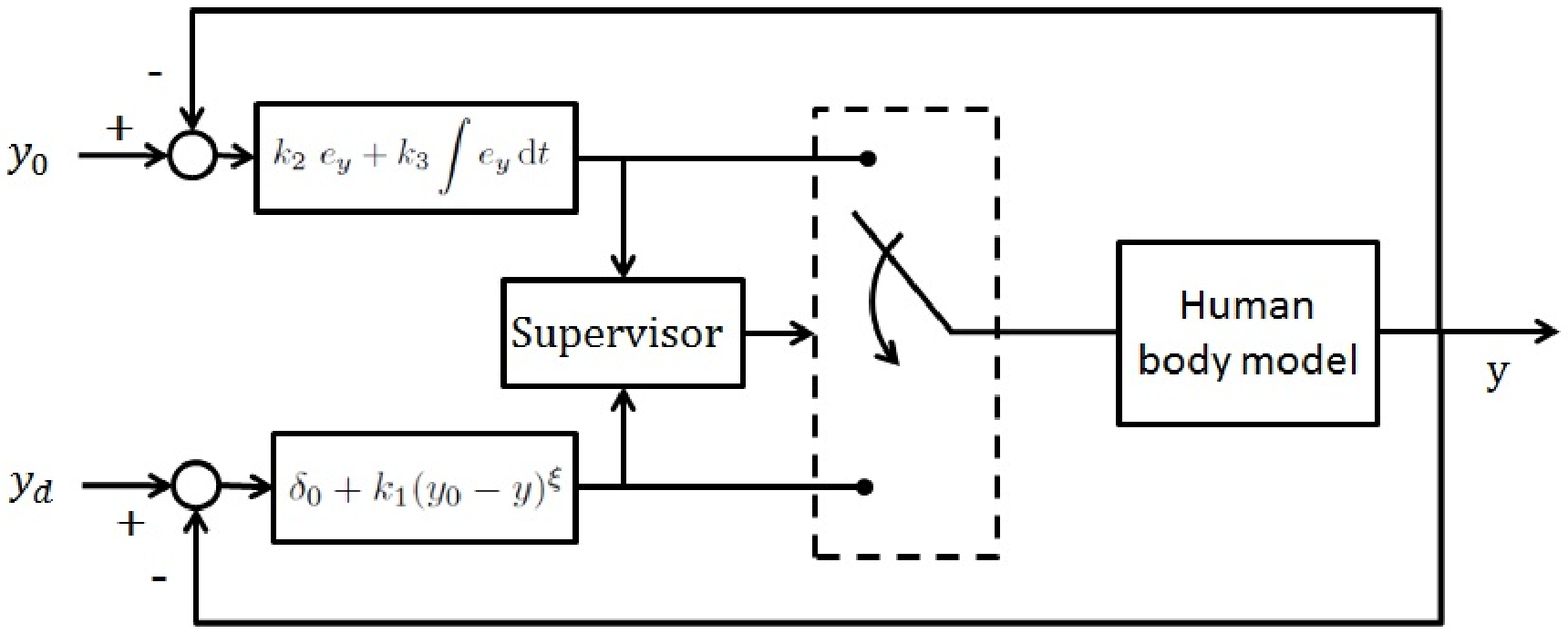}
}
\end{center}
\caption{Controller block diagram: (a) time delayed and (b) switching-based}
\label{fig:block}
\end{figure}

The linearizing feedback control law results in
\begin{equation} \label{eq:u}
u =  \bar{g}^{-1} (\dot{y_d}-H(t)-\Pi e_y)
\end{equation}
Note in equation \eqref{eq:u} that the time delayed estimation of the total sum of system nonlinearities is used in place of the total sum of system nonlinearities. Namely,
\begin{equation} \label{eq:H}
H(t) \approx H(t-\Delta) = \dot{y}(t-\Delta)-\bar{g} u(t-\Delta)
\end{equation}
Combining equations \eqref{eq:u} and \eqref{eq:H}, the time delayed control law (Fig. \ref{fig:block}(a)) can be obtained as
\begin{align} \label{eq:tdc}
u = \delta_{TDC} &=  \bar{g}^{-1} (\dot{y_d}-H(t-\Delta)-\Pi e_y)  \nonumber \\
&= u(t-\Delta) + \bar{g}^{-1}(\dot{y_d}-\dot{y}(t-\Delta)-\Pi e_y)
\end{align}
The stability condition was shown in \cite{Youcef-Toumi--Wu1992}.
If a specific trajectory for weight loss is desired then, we can address the tracking problem. In this controller, not only the current and previous outputs are used, but the previous input is also considered, which could make it more efficient for tracking purposes.

In many circumstances, a modest initial increase in physical activity is preferred at the beginning of the program. Switching-based set-point regulation is an alternative way to move the body weight to a desired value considering the predefined  physical activity shape. A possible formula for $\delta$ is given by
\begin{equation}
u = \delta_{INC}=\delta_0+k_1(y_0-y)^{\xi}
\end{equation}
The parameters $\delta_0$ and $y_0$ are the initial values for the $\delta$ and $y$, respectively. To steer the human body to a particular weight, the following algorithm is proposed (see Fig. \ref{fig:block}(b)): 
\begin{align*}
&\delta=\delta_{INC} \\ 
&\text{IF}~~\delta_{INC}>\delta_{PI} \\
&~~~~~\delta=\delta_{PI} \\
&\text{END},
\end{align*}
where $\delta_{PI}$ is a proportional-integral controller as
\begin{equation}
u = \delta_{PI} = k_2~e_y + k_3 \int e_y \, \mathrm{d}t  
\end{equation}
The control parameters are chosen in such a way that $\delta_{PI}$ is greater than $\delta_{INC}$ at the beginning. So, the control action is started with $\delta_{INC}$ and then switched to $\delta_{PI}$ when it is less than $\delta_{INC}$.  

\vspace{1mm}
\section*{Observer Design} \label{sec:Observer}

The state vector containing fat, lean and extra cellular fluid masses is not easy to measure directly every day. Therefore, because of the cost and inconvenience, it is desirable to estimate these from the system input and body weight.

The observer is designed to provide real-time estimates of inaccessible dynamical states that might be required for the implementation of control laws.

Substituting $r$ from~\eqref{eq:r} into~\eqref{eq:Fat} and~\eqref{eq:Lean} leads to 
\begin{equation} \label{eq:fat_r_subs}
\rho_{F} \frac{dF}{dt}=(1-\frac{1}{1+\frac{F}{c}})EB
\end{equation}
\begin{equation} \label{eq:lean_r_subs}
\rho_{L} \frac{dL}{dt}=\frac{1}{1+\frac{F}{c}}EB
\end{equation}

To cope with the non-linearity, the following linear approximation is made
\begin{equation} \label{eq:regression}
\frac{1}{1+\frac{F}{c}} \approx a+bF
\end{equation}
Constants $a$ and $b$ must be determined based on the regression. Then, Eqs.~\eqref{eq:fat_r_subs} and~\eqref{eq:lean_r_subs} are rewritten regarding the approximation as well as expansion of the $EB$ components obtained from Eq. \eqref{eq:EB}.

\begin{equation} \label{eq:fat_reg_subs}
\small \rho_{F} \frac{dF}{dt}=(1-a-bF)(EI - [\delta \cdot BM] - [\beta \cdot EI] - \left[ K+\gamma_L L+ \gamma_F F \right])
\end{equation}
\begin{equation} \label{eq:lean_reg_subs}
\small \rho_{L} \frac{dL}{dt}=(a+bF)(EI - [\delta \cdot BM] - [\beta \cdot EI] - \left[ K+\gamma_L L+ \gamma_F F \right] )
\end{equation}

For observer design, we arrange the obtained system of differential equations of \eqref{eq:fat_reg_subs}, \eqref{eq:lean_reg_subs} and \eqref{eq:ECF} as the following class of nonlinear systems:
\begin{equation} \label{ABPhi}
\dot{x} = h(t,x,y,u) = Ax + Bu + \Phi(t,x,y,u)
\end{equation}
\begin{equation} \label{y-ABPhi}
y = Cx
\end{equation}
where $x=[F,~L,~ECF]$ and $A$ is a constant matrix.

It can be shown that $\Phi(t,x,y,u)$ is Lipschitz i.e. 
\begin{equation}
\| \Phi-\hat{\Phi} \| \leq \lambda \|(x-\hat{x})\|
\end{equation}
and designing an observer for Eq.~\eqref{ABPhi} is feasible if the pair (A,C) is observable.

\begin{equation}
A=
\begin{bmatrix}
\frac{-(1-a)\gamma_F+bK}{\rho_{F}} & -\frac{(1-a)\gamma_L}{\rho_{F}} & 0 \\
\frac{-a\gamma_F-bK}{\rho_{L}} & \frac{-a\gamma_L}{\rho_{L}} & 0 \\
0 & 0 & -\frac{\rho_{w}}{Na} \zeta_{Na}
\end{bmatrix}
\end{equation}

\begin{equation}
C = [1~1~1]
\end{equation}

The Luenberger observer is well-known and widely used for time-invariant linear systems in order to estimate the state vector when that is not directly measurable. Consider the following extended Luenberger observer for the system \eqref{ABPhi}-\eqref{y-ABPhi} \cite{Pagilla--Zhu2004}:
\begin{equation} \label{eq:observer1}
\dot{\hat{x}} = A \hat{x}+Bu+\Phi(t,\hat{x},y,u)+\frac{\lambda^2+\epsilon_0}{||C||^2}G(y-C\hat{x})+G_1(y-C\hat{x})
\end{equation}
where $\epsilon_0 \geq -\lambda^2$ and $G_1$ is chosen such that $A-G_1C$ is Hurwitz, and $G$ is the observer gain matrix. Let the estimation error 
\begin{equation}
\tilde{x} = x-\hat{x}  
\end{equation}
in which $P_0$ is positive definite matrix. Then, by considering the following Lyapunov function candidate
\begin{equation}
V(\tilde{x}) = \tilde{x}^T P_0  \tilde{x}  
\end{equation}
the observer gain $G$ is obtained as
\begin{equation}
G=P^{-1}_0 C^T /2
\end{equation}

\begin{figure}
  \centering
  \includegraphics[trim = 0in 0in 0in 0in, clip, scale=.4]{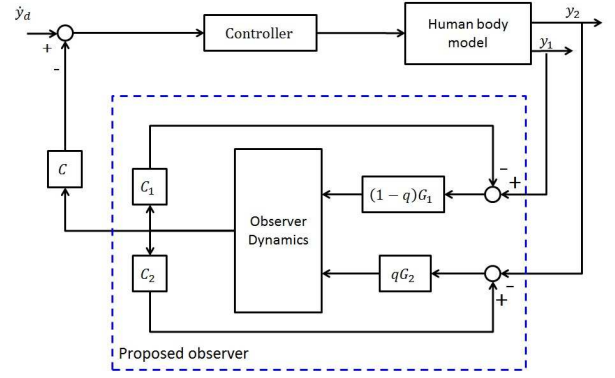}
  \caption{Proposed observer block diagram}
  \label{fig:obs}
\end{figure}

In reality, it may not be possible to measure precisely both energy intake and energy expenditure every day. To overcome this problem, it is assumed that the measurement of body components is performed periodically. Hence, in Eq. \eqref{eq:observer1} with $\epsilon_0 = -\lambda^2$ the following soft switching observer is presented (see Fig. \ref{fig:obs})
\begin{equation} \label{eq:observer2}
\small \dot{\hat{x}} = A \hat{x}+Bu+\Phi(t,\hat{x},y,u)
+(1-q)G_1(y_1-C_1\hat{x}) +q G_2(y_2(jT)-C_2\hat{x})
\end{equation}

\begin{equation} \label{eq:observer2}
 \dot{\hat{x}} = f(\hat{x},y,u)+g(\hat{x},y)u
+(1-q)G_1(y_1-C_1\hat{x}) +q G_2(y_2(jT)-C_2\hat{x})
\end{equation}

\begin{equation}
y_1 = C_1x,~~y_2(jT)=C_2x(jT),~~j=1,2,...
\end{equation}
in which $C_1=[1~1~1]$, $C_2=1$, $G_i$s are observer gain matrices, $T$ is the period that measurement of components is performed, and $0\leq q \leq 1$ is a weighting factor for soft switching:  
\begin{equation}
q = exp(-\frac{t-jT}{k_q}),~~~~~jT \leq t<(j+1)T
\end{equation}
$k_q$ is appropriately determined.

Using intermittent measurement of full states in addition to daily body weight measurement, the degree of observability would be increased.

\section*{Simulation Results} \label{sec:Simulation}

In this section, we present simulation results based on our proposed controllers and observers. We use the following coefficients and parameters of the dynamic system \cite{Sentocnik--AtanasijevicKunc--Drinovec--Pfeifer2013}.
\vspace{2mm}
\\
$k_1=4~kcal/g$, $k_2=9~kcal/g$, $k_3=4~kcal/g$, $\rho_F = 9400~kcal (kg/d)$, $\rho_L = 1800~kcal/(kg/d)$, $kk = 10.4~kg$, $\rho_{w} = 1~kg/l$, $Na = 3.22~kg/l$, $\zeta_{Na} = 3~kg/d/l$, $\zeta_{ci} = 4~kg/d$, $\beta = 0.24$, $\gamma_L = 22~kcal/kg$, $\eta_L = 230~kcal/(kg/d)$, $\gamma_F = 3.2~kcal/kg$, $\eta_F = 180~kcal/(kg/d)$.
\vspace{2mm}
\\
The control objective in each simulation is tracking or regulation of body or fat masses with initial conditions $[F,~L,~ECF,~\hat{F},~\hat{L},~\hat{ECF}]=[30,~45,~25,~ 31,~44,~26]$. 

\begin{figure}
  \centering
  \includegraphics[trim = 0in 0in 0in 0in, clip, scale=.5]{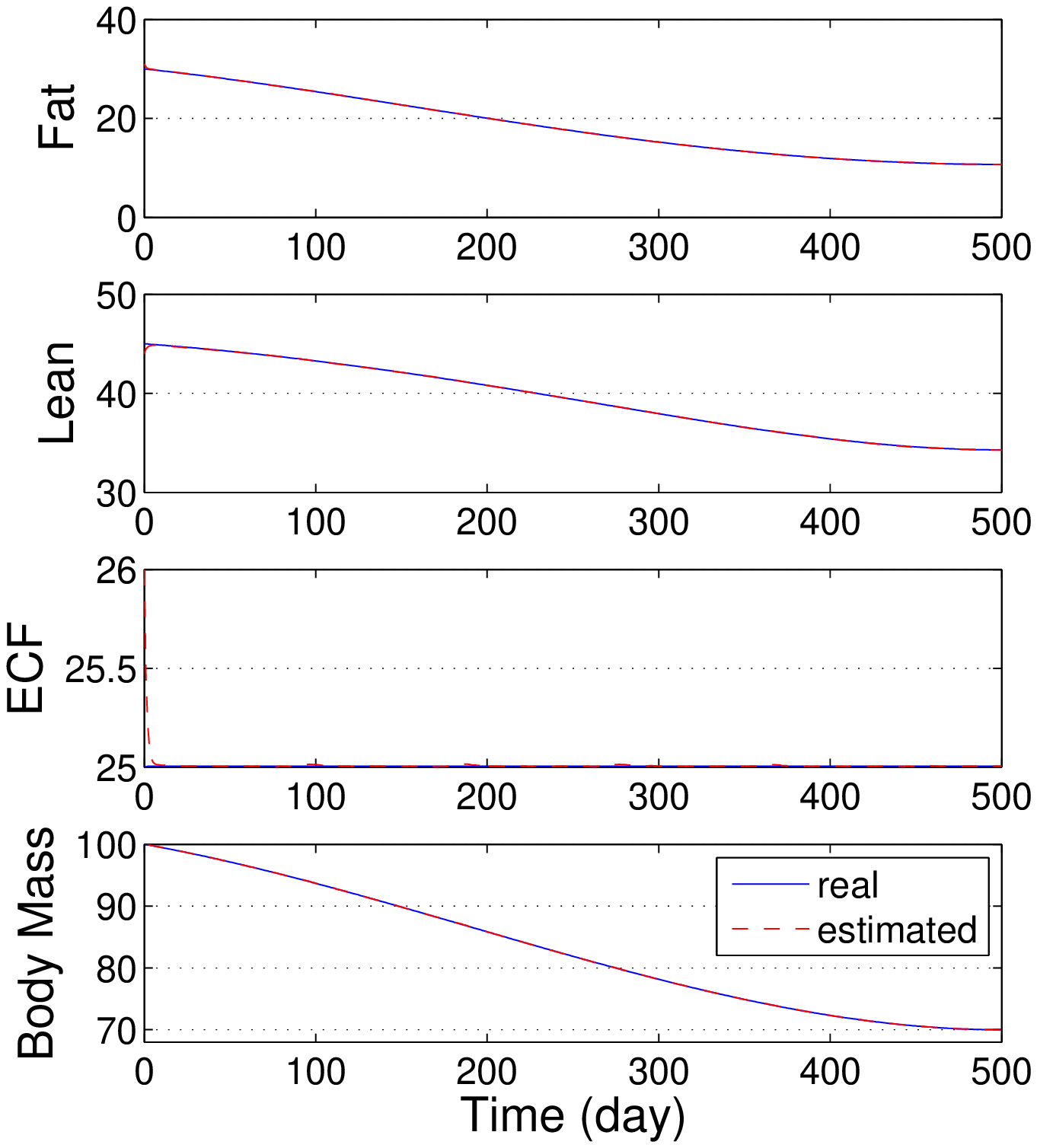}
  \caption{The fat (F), lean (L) and extracellular fluid (ECF): time delayed body mass controller}
  \label{fig:massC1bm}
\end{figure}

\begin{figure}
  \centering
  \includegraphics[trim = 0in 0in 0in 0in, clip, scale=.4]{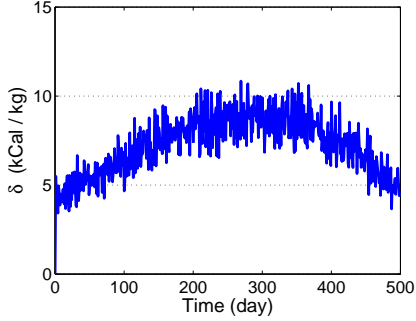}
  \caption{Control input $\delta$: time-delayed body mass controller}
  \label{fig:delC1bm}
\end{figure}

The influence of $EI=3492(1+0.02*rand),~~ \delta=\delta(1-0.2*rand)$ disturbances on controllers and the soft switching observer are investigated in the following subsections. The variable $rand$ is a MATLAB function that generates uniformly distributed pseudo-random numbers between 0 and 1. The coefficient of 0.2 for disturbance in the physical activity could be up to 200~kcal of energy expenditure in this case. Since $EI$ is multiple times greater than physical activity ($\delta \cdot BM$), the coefficient of $rand$ in $EI$ is chosen much less frequently than the coefficient of $rand$ in $\delta$.

For the observer \eqref{eq:observer2}, measurement of the body components is done every 90 days with the following parameters hereafter: \\
$G_1=[1.9~ 1.2~ -0.3],~~G_{2ii}=[0.8~ 1~ -0.2],~~k_q=5$ 
\subsection*{Body mass controller}
Next, we address how to control body weight according to a desired cubic polynomial function using the controller \eqref{eq:tdc} even in the presence of disturbances. The initial and final values of the body weight are 100 kg with slope -0.05 and 70 kg with slope 0, respectively. The control parameters considered in this simulation are $\bar{g}=0.1$ and $\Pi=10$. Since the controlled output is body weight and it is measurable directly, controlling based on observer is not needed.

The results of the body weight control are shown in Figs. \ref{fig:massC1bm}--\ref{fig:delC1bm}. The lean, fat and ECF masses are bounded without fluctuation while body weight is being controlled (Fig. \ref{fig:massC1bm}). The controller error is small in the presence of disturbances, which implies good performance of the controller. The control input $\delta$ in Fig. \ref{fig:delC1bm} has a dome shape using a time-delayed controller for a cubic polynomial trajectory.

\begin{figure}
  \centering
  \includegraphics[trim = 0in 0in 0in 0in, clip, scale=.5]{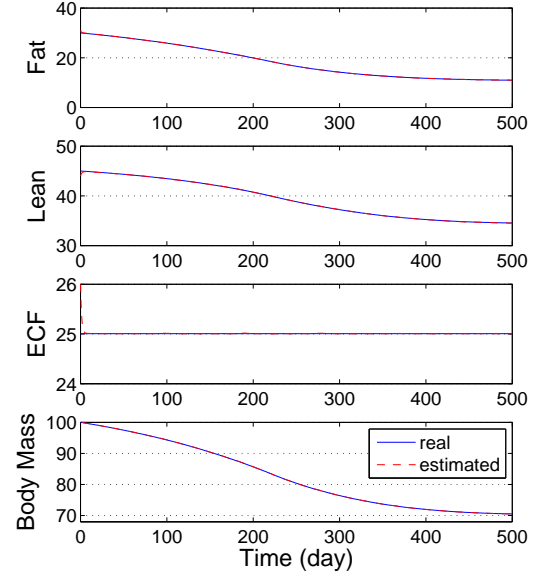}
  \caption{The fat (F), lean (L) and extracellular fluid (ECF): switching-based body mass controller}
  \label{fig:massC2bm}
\end{figure}

\begin{figure}
  \centering
  \includegraphics[trim = 0in 0in 0in 0in, clip, scale=.4]{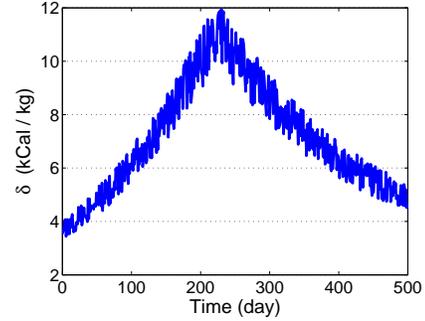}
  \caption{Control input $\delta$: switching-based body mass controller}
  \label{fig:delC2bm}
\end{figure}

The results of the switching-based controller are presented in Figs. \ref{fig:massC2bm} through \ref{fig:delC2bm}. The control parameters chosen for this controller are:\\
$\delta_0=4,~~ k_1=0.5,~~ \xi=1,~~ k_2=0.2,~~k_3=0.4$

It is seen that the desired body weight is achieved even in the presence of disturbances. The control input $\delta$ in Fig. \ref{fig:delC2bm} has a hat $(~\hat{}~)$ shape that is different from the dome shape of the time-delayed controller. We find that although the maximum of the dome shape is lower than the maximum of the hat shaped exercise profile, the desired weight can be achieved.

~~~\subsection*{Fat mass controller}
\begin{figure}
  \centering
  \includegraphics[trim = 0in 0in 0in 0in, clip, scale=.5]{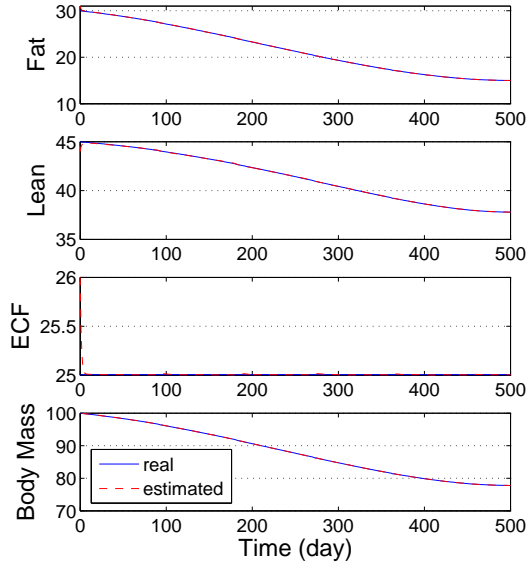}
  \caption{The fat (F), lean (L) and extracellular fluid (ECF): time delayed fat mass controller}
  \label{fig:massC1fat}
\end{figure}

\begin{figure}
  \centering
  \includegraphics[trim = 0in 0in 0in 0in, clip, scale=.4]{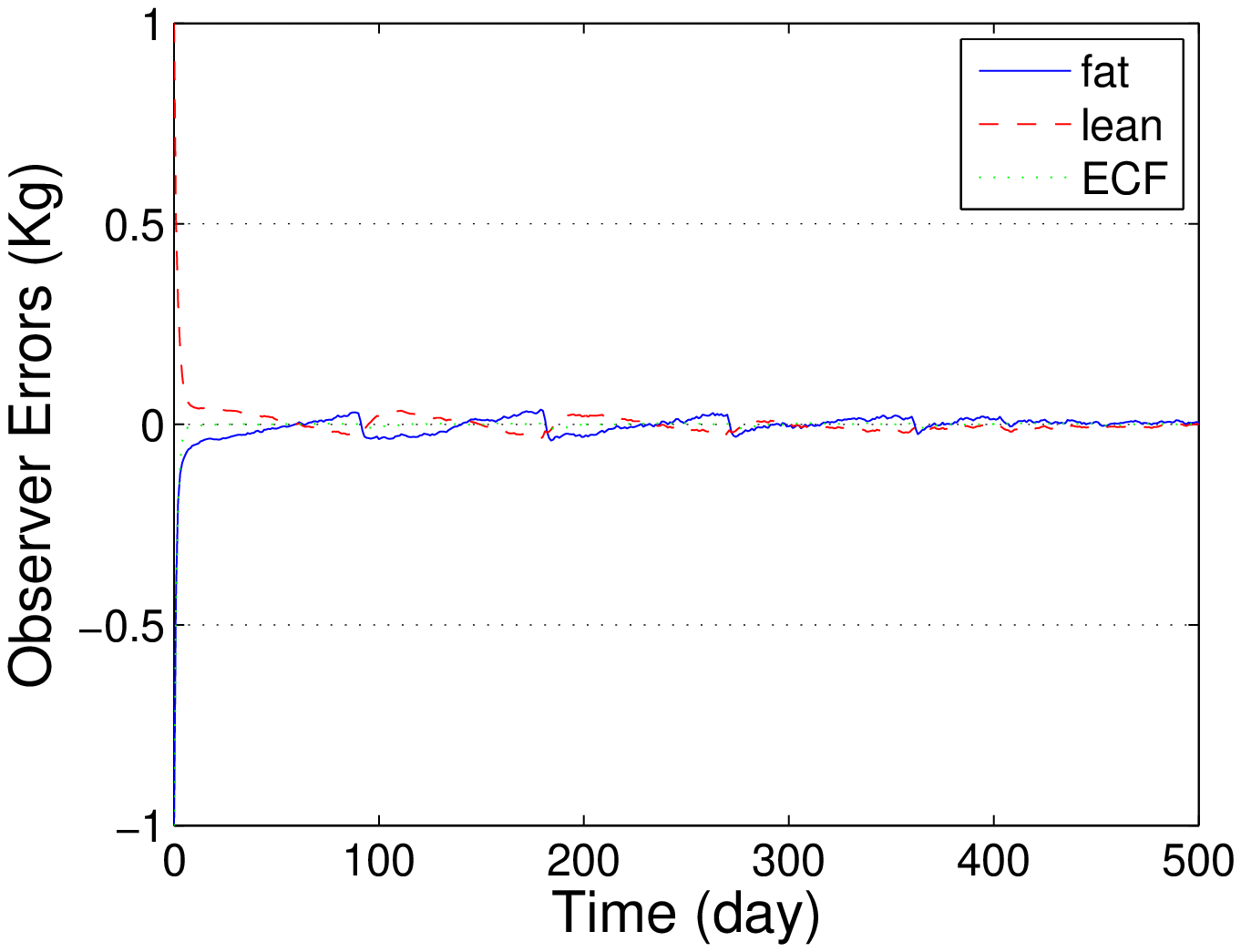}
  \caption{Observer errors: Eq. \eqref{eq:observer2}}
  \label{fig:OerC1fat}
\end{figure}

\begin{figure}[]
  \centering
  \includegraphics[trim = 0in 0in 0in 0in, clip, scale=.4]{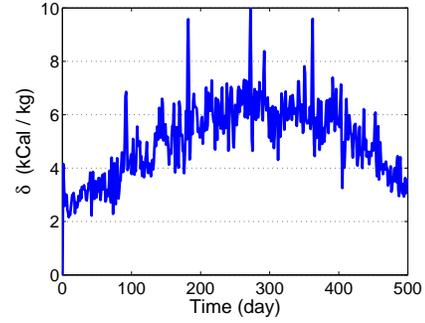}
  \caption{Control input $\delta$: time delayed fat mass controller}
  \label{fig:delC1fat}
\end{figure}

Here, the objective is to control fat mass to bring it to a desired cubic polynomial function using the time-delayed controller. The initial and final values of the fat mass are 30 kg with slope -0.02 and 15 kg with slope 0, respectively. All controller and observer parameters are the same as the 1st simulation for body mass control. Since the controlled output is fat mass and direct measurements are only available periodically, observer-based control is necessary. Simulation results of the controller and observer \eqref{eq:observer2} are represented in Figs \ref{fig:massC1fat}--\ref{fig:delC1fat}. In addition to the decreasing and bounded body composition masses in Fig. \ref{fig:massC1fat}, small observer errors in Fig. \ref{fig:OerC1fat} are achieved. There are some sharp changes in fat mass error and spikes in $\delta$ (see Fig. \ref{fig:delC1fat}), which come from rapid observer corrections due to periodic measurement.

\begin{figure}[]
  \centering
  \includegraphics[trim = 0in 0in 0in 0in, clip, scale=.5]{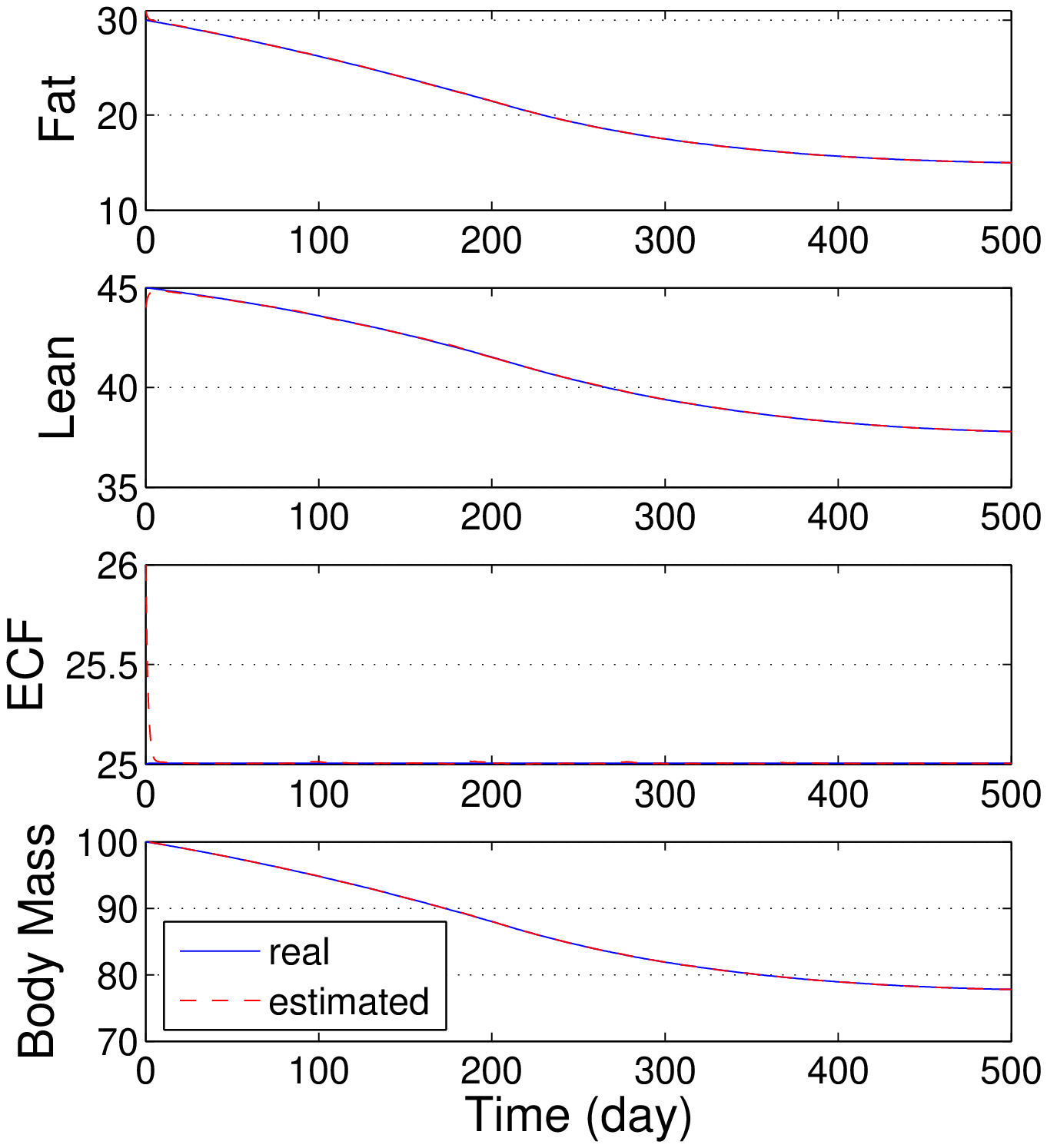}
  \caption{The fat (F), lean (L) and extracellular fluid (ECF): switching-based fat mass controller}
  \label{fig:massC2fat}
\end{figure}

\begin{figure}[]
  \centering
  \includegraphics[trim = 0in 0in 0in 0in, clip, scale=.4]{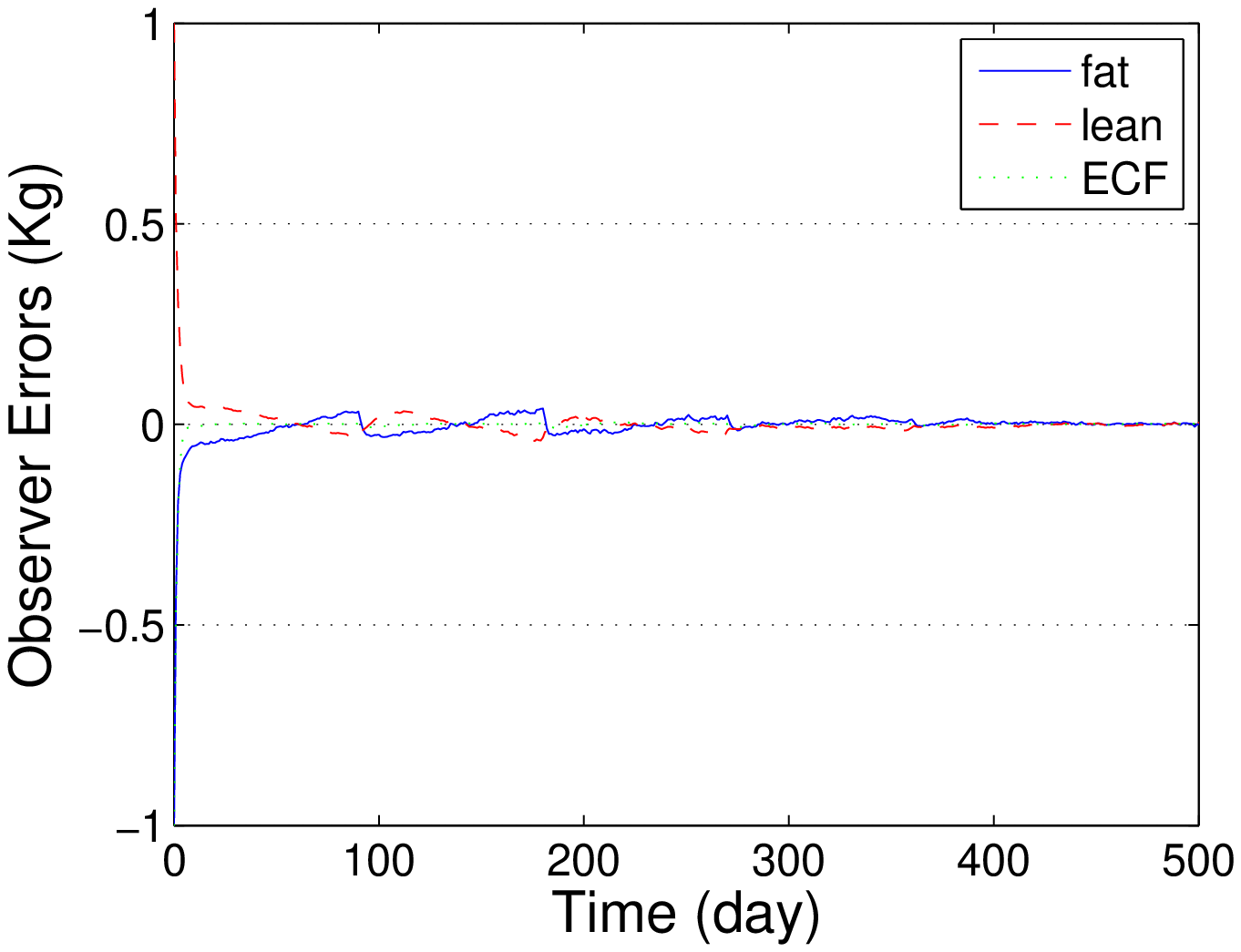}
  \caption{Observer errors: Eq. \eqref{eq:observer2}}
  \label{fig:OerC2fat}
\end{figure}

\begin{figure}[]
  \centering
  \includegraphics[trim = 0in 0in 0in 0in, clip, scale=.4]{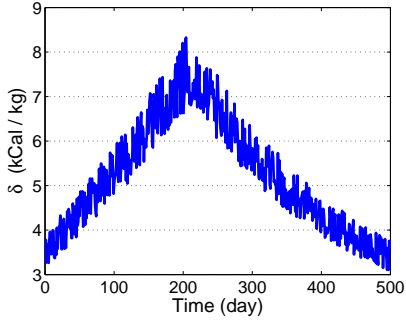}
  \caption{Control input $\delta$: switching-based fat mass controller}
  \label{fig:delC2fat}
\end{figure}

The last simulation shows the results of a switching-based controller for the regulation of fat mass using the observer of \eqref{eq:observer2} depicted in Figs. \ref{fig:massC2fat}--\ref{fig:delC2fat}. The initial and final values of the fat mass are 30 kg and 15 kg, respectively. All parameters are similar to the parameters of the same controller for body mass, but $k_2=0.7$.

In summary, observer-based time delayed control methods perform well in the trajectory tracking problem as shown in the results of the fat mass controller where at least one of the body component measurements is needed. In the case of total body mass control, the observer is not necessary in the time delayed controller because the body weight is directly measurable. Also, a switching-based controller works well for set-point regulation, and the same discussion is valid for requiring an observer for this controller.

\section{CONTROL ALLOCATION}

\vspace{1mm}
\section*{Constrained Optimal Control Design} \label{sec:controller}

Let the control objective be steering any combination of body composition masses, such as body mass, to a desired trajectory. It is assumed that both energy intake and physical activity are the control inputs to the system. This shows overactuation and there are an infinite number of solutions. Hence, a control allocation procedure is required to find the best solution that fits within the objective and limitations of the problem. 

The body weight dynamics~\eqref{eq:Fat}-\eqref{eq:ECF} can be formed as
\begin{equation} \label{eq:fgu}
\dot{x} = h(t,x,y,u) = f(t,x,y)+g(x,y)u
\end{equation}
\begin{equation} \label{eq:yallo}
y=Cx
\end{equation}

in which $x=[F,~L,~ECF]$, $u=[EI,~\delta]$, $y$ is a controlled output, $C$ is a $1 \times 3$ matrix and
\begin{equation} \label{eq:f}
\small f(t,x,y)=
\begin{bmatrix}
- (1-a-bF) \left[ K+\gamma_L L+ \gamma_F F \right]) \\
- (a+b F) \left[ K+\gamma_L L+ \gamma_F F \right] ) \\
\frac{\rho_{w}}{Na} \left[ 
\zeta_{Na}(ECF_{init}-ECF) - \zeta_{ci}
\right]
\end{bmatrix}
\end{equation}

\begin{equation} \label{eq:g}
\small
g(x,y)=
\begin{bmatrix}  
(1-\beta)(1-a-b F) &  -(1-a-b F)BM \\
(1-\beta)(a+b F)   &  -(a+b F)BM \\
\frac{\rho_{w}}{Na} \zeta_{ci}\frac{z_1}{ci_b}   &   0
\end{bmatrix}
\end{equation}
It is presumed that the carbohydrate energy intake is proportional with total energy intake.
\begin{equation}
k_1 ci = z_1 EI
\end{equation}

Differentiating the output \eqref{eq:yallo} with respect to $t$ yields
\begin{equation} \label{eq:ydotALLO}
\dot{y} = Cf(x,y)+Cg(x,y)u
\end{equation}

Let the desired dynamics for the feedback linearization be defined as
\begin{equation} \label{eq:errDyn}
\dot{e}_y + \lambda_{fl} e_y = 0 ~~\text{or}~~  \dot{y} = \dot{y}_{des} - \lambda_{fl} e_y
\end{equation}
Then, substituting \eqref{eq:errDyn} into \eqref{eq:ydotALLO} results in
\begin{equation} 
[Cg(x,y)]~u = [\dot{y}_{des} - \lambda_{fl} e_y -Cf(x,y)]
\end{equation}

For sliding mode control (SMC), instead of setting the design surface temperature as the output, the sliding surfaces $S=e_y+\lambda_{sm} \int e_y \, \mathrm{d}t$ are selected as the outputs. With only one differentiation 

\begin{align}
\dot{S} &= \dot{e}_y + \lambda_{sm} e_y = \dot{y} - \dot{y}_{des} + \lambda_{sm} e_y \nonumber \\
&= Cf(x,y) + Cg(x,y)u - \dot{y}_{des} + \lambda_{sm} e_y \\
&= - \Gamma  \cdot sgn(S) \nonumber
\end{align}

\begin{equation} 
[Cg(x,y)]~u = [\dot{y}_{des} - \lambda_{sm} e_y -Cf(x,y) - \Gamma  \cdot sgn(S)]
\end{equation}

Now approaching the system as a control allocation problem, we define the optimization problem as
\begin{align} \label{eq:OptCont}
u_s = &\arg \min_u(\frac{1}{2} (u-\bar{u})^T W (u-\bar{u}))~~\text{subject to} \nonumber \\
&Au=b \\
&u_{i \hspace{.5mm} min} \leq u \leq u_{i \hspace{.5mm} max}, ~~i=1,2 \nonumber
\end{align}
where $A=Cg(x,y)$
\begin{equation}
 \small b =
  \begin{cases}
   \dot{y}_{des} - \lambda_{fl} e_y -Cf(x,y)       & \text{for } FBL \\
   \dot{y}_{des} - \lambda_{sm} e_y -Cf(x,y) - \Gamma  \cdot sgn(S)       & \text{for } SMC
  \end{cases}
\end{equation}
The objective function in \eqref{eq:OptCont} is given by the quadratic function of the control inputs in which $W$ is a weighting matrix. 

Since $A$ is a full row rank matrix and the dimension of $u$ is larger than the number of equations, the problem is redundant and has an infinite number of solutions.

Now, the aforementioned control allocation problem with input limitation is solved using the fmincon optimization solver in MATLAB. The $u_{i\hspace{.5mm}min}$ and $u_{i \hspace{.5mm} max}$ should be within bounds so that the allocation problem becomes feasible. The threshold values defining the bounds could depend on various factors, such as physical limits and initial and boundary conditions.

Logically, to stop an increase in body mass, the maximum allowable amount of energy intake would be $\bar{EI}$ at the beginning of the program. It means that the first component of $u_s=[u_{s1}~~~u_{s2}]$ should be equal to or less than $\bar{EI}$ at the initial time that is $u_{s1}(t=0)\leq \bar{EI}$. Hence, the minimum value for energy intake changes over time and is proposed as 
\begin{equation}
u_{1\hspace{.5mm} min}=\bar{EI}- \rho_1(1- \rho_2~ exp(-t/\tau)),~~u_{1 \hspace{.5mm} max}=\bar{EI}
\end{equation}
where the parameters $\rho_1$, $\rho_2$ and $\tau$ are for determining lower bound, upper bound and speed of convergence from upper bound to lower bound of $u_{s1}$ to design appropriate behavior. At the beginning of treatment, $EI_{min}$ is equal to $\bar{EI}- \rho_1(1- \rho_2)$ and after the passing some time, it is $\bar{EI}- \rho_1$. 

\vspace{1mm}
\section*{Simulation Results} \label{sec:Simulation}
In this section, we consider simulations that rely on the proposed allocated controller for comparative analysis and performance investigation. The control objective in each simulation is computing energy intake and physical activity coefficients using \eqref{eq:OptCont} to track a desired cubic polynomial function of total body mass in the presence of uncertainty, noise and disturbances. The initial and final values of the body weight are 100 kg with slope -0.05 and 70 kg with slope 0, respectively. The control parameters considered in this simulation are \\
$\lambda_{fl}=0.7,~~ \lambda_{sm}=0.1,~~W_{11}=0.2,~~W_{22}=1000,~~W_{ij}=0,~~ \Gamma=0.01,~~\bar{u}=[3500~~0],~~\rho_2=1,~~\tau=5,~~\bar{EI}=3500$\\
For true comparison, the bound of tracking errors for both controllers are considered to be equal so that $\lambda_{sm}$ would be 7 times greater than $\lambda_{fl}$. The applied uncertainty to the system and noise to the measured daily body weight are $r = (1+0.5*rand)*r$ and $bm = bm+0.1*rand$, respectively and $EI=3492(1+0.01*rand)$ and $\delta=\delta(1-0.1*rand)$ disturbances inserted in to the inputs. The $rand$ is a MATLAB function that generates uniformly distributed pseudo-random numbers between 0 and 1.

\begin{figure}
\begin{center}
\subfigure[]{
\small
\includegraphics[trim = 0in 0in .4in 0in, clip, scale=0.3]{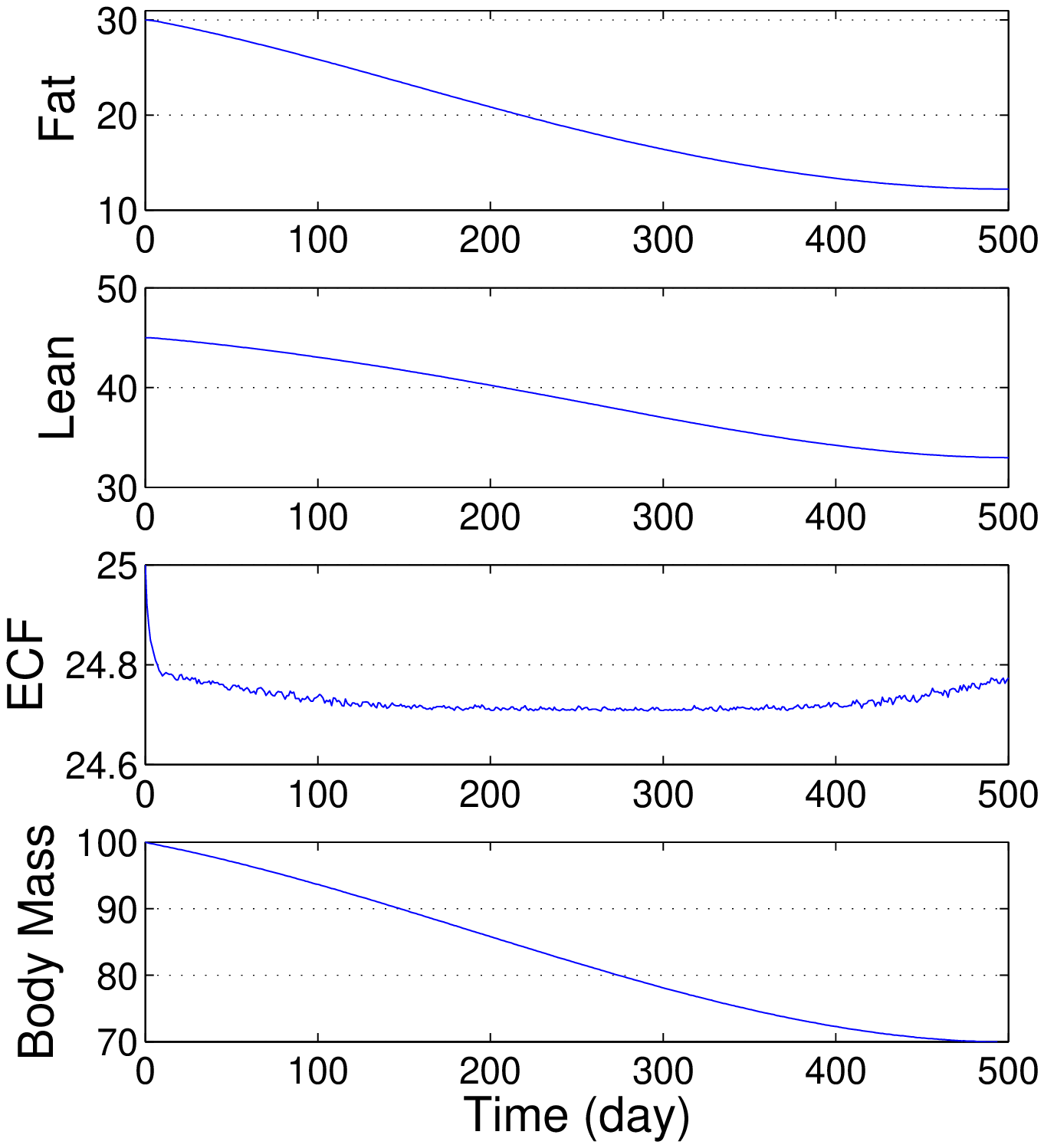}
}
\subfigure[]{
\small
\includegraphics[trim = 0in 0in .4in 0in, clip, scale=0.3]{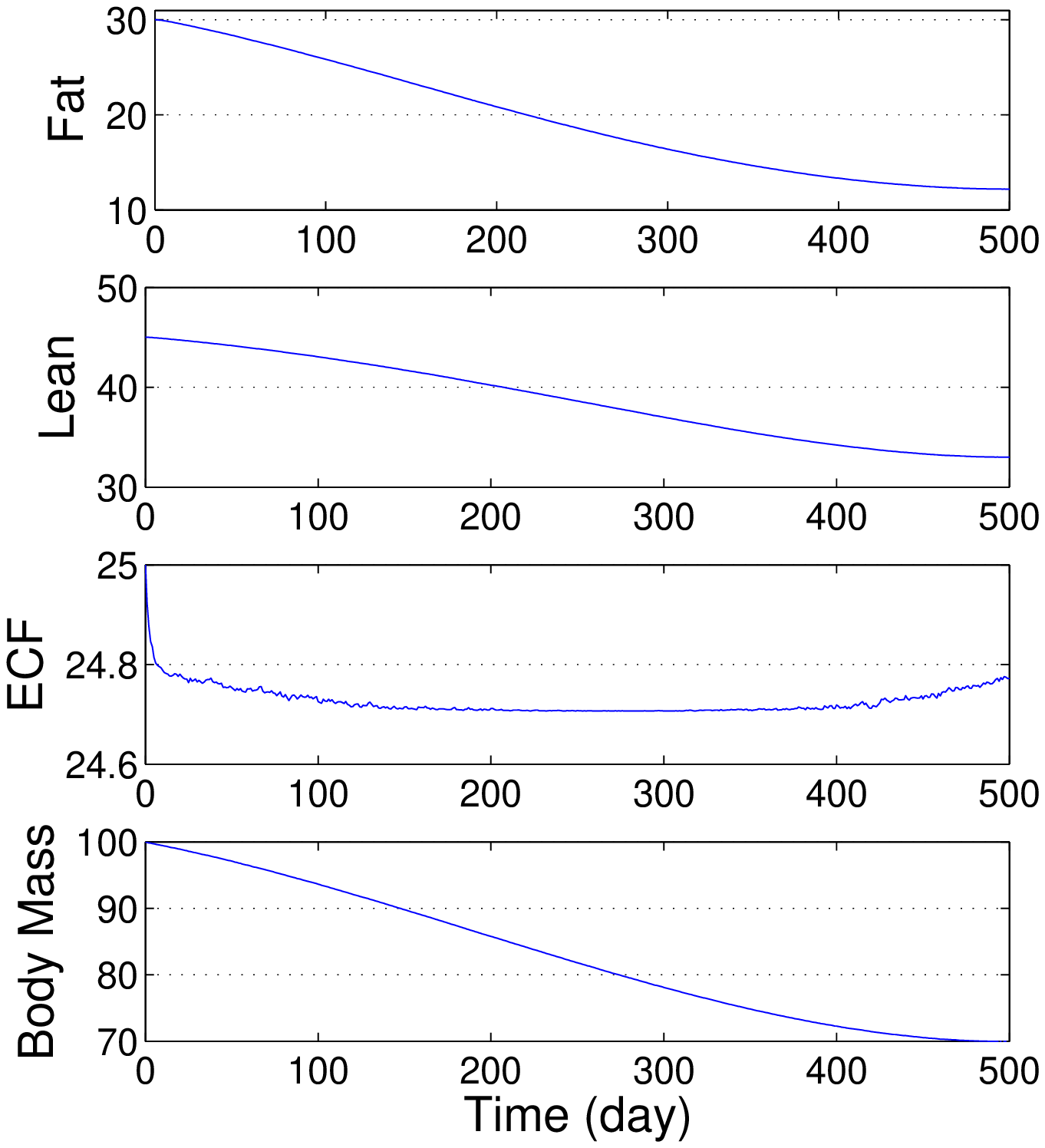}
}
\end{center}
\caption{The fat (F), lean (L) and extracellular fluid (ECF) for $\rho_1=500$: (a) FL and (b) SMC}
\label{fig:mass2}
\end{figure}

\begin{figure}
\begin{center}
\subfigure[]{
\small
\includegraphics[trim = 0in 0in .4in 0in, clip, scale=0.29]{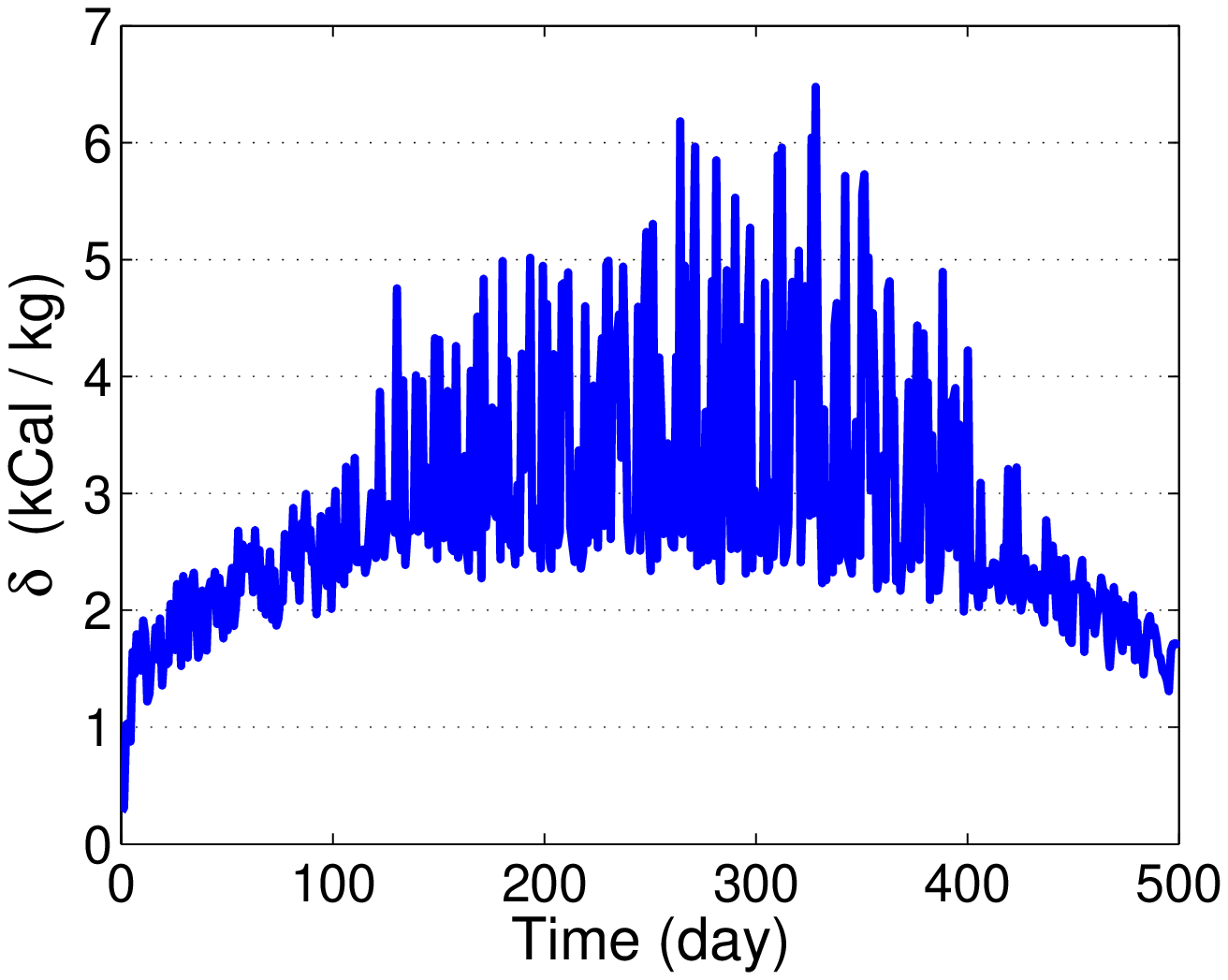}
}
\subfigure[]{
\small
\includegraphics[trim = 0in 0in .4in 0in, clip, scale=0.3]{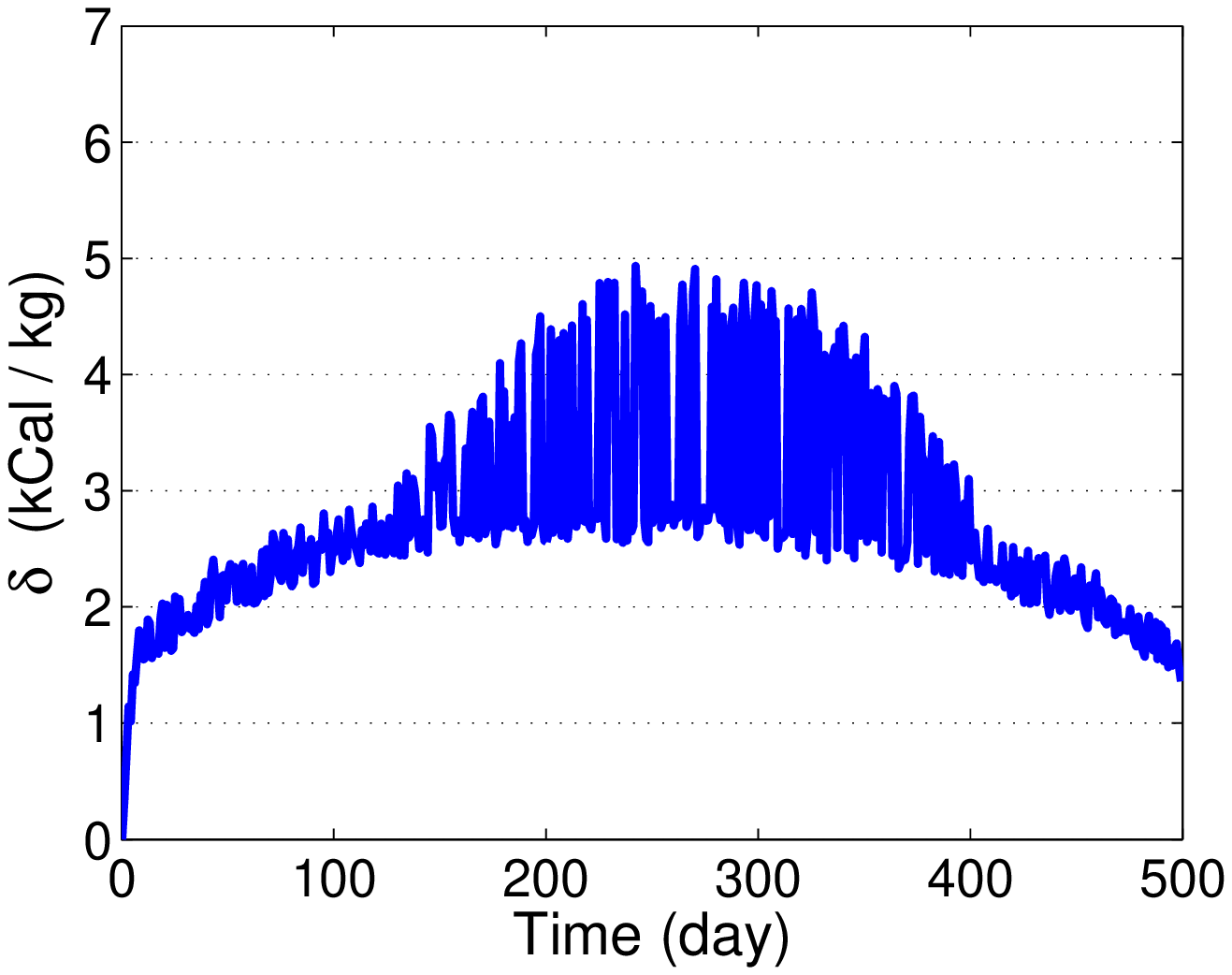}
}
\end{center}
\caption{Physical Activity Coefficient ($\delta$) for $\rho_1=500$: (a) FL and (b) SMC}
\label{fig:del2}
\end{figure}

\begin{figure}
\begin{center}
\subfigure[]{
\small
\includegraphics[trim = 0in 0in .4in 0in, clip, scale=0.3]{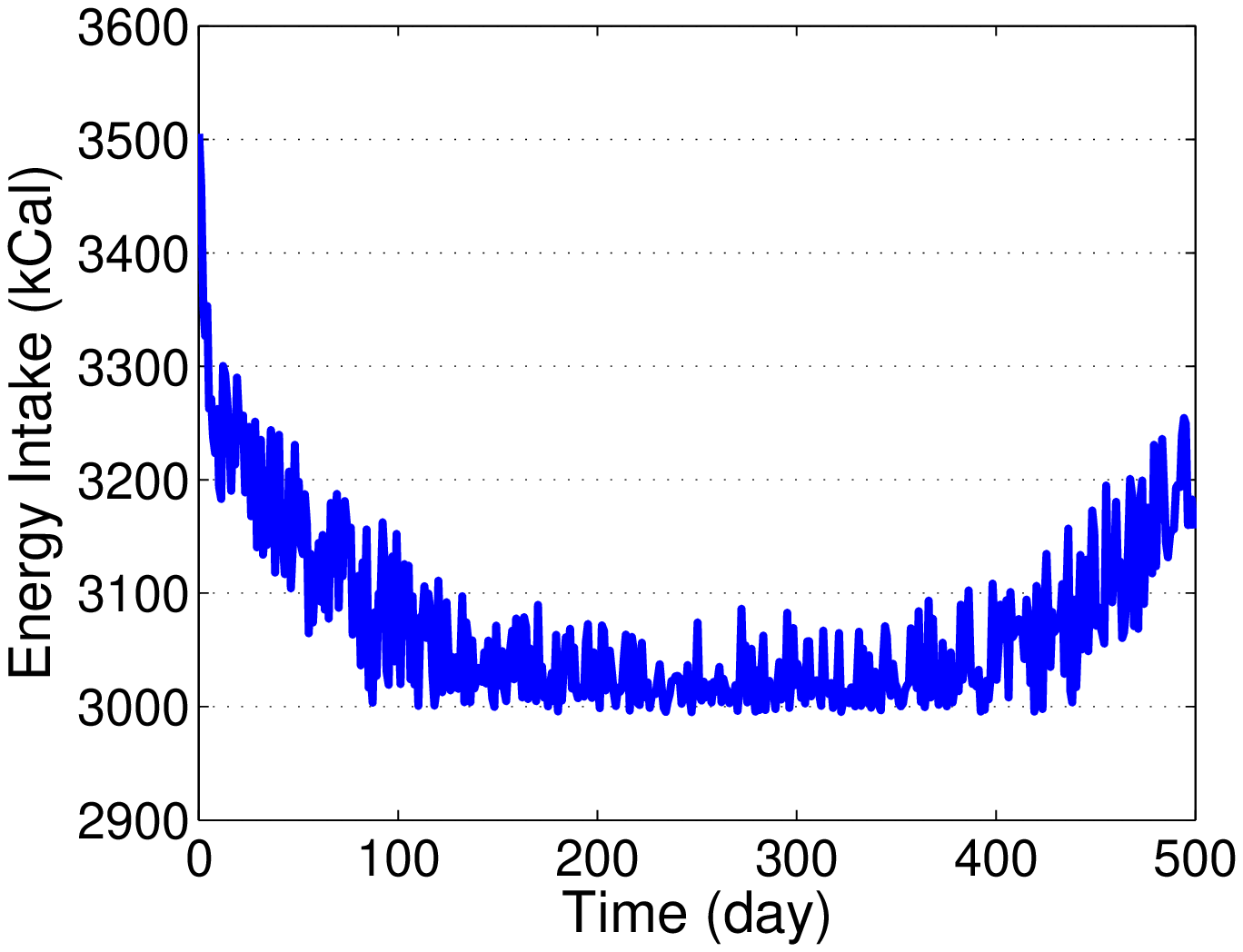}
}
\subfigure[]{
\small
\includegraphics[trim = 0in 0in .4in 0in, clip, scale=0.3]{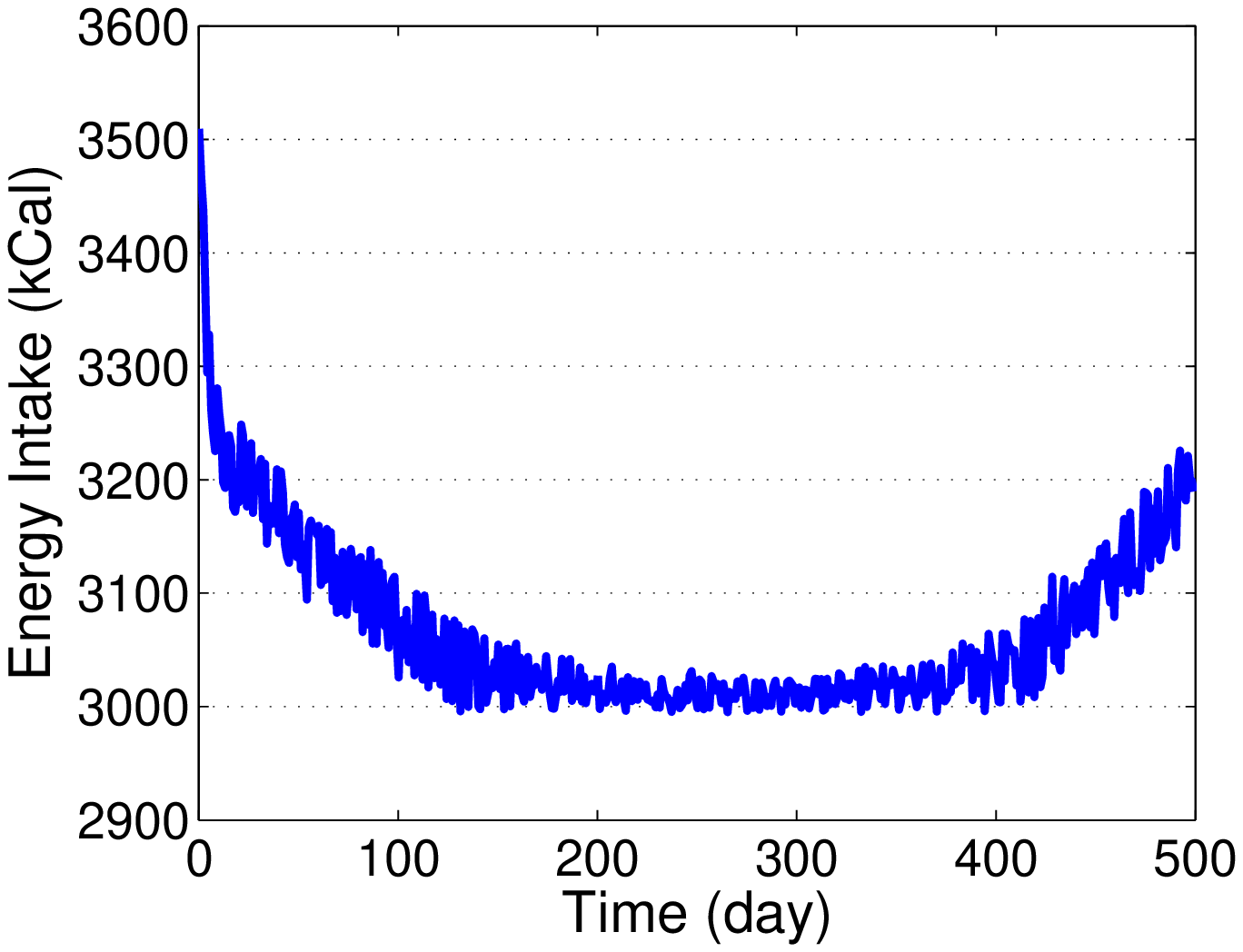}
}
\end{center}
\caption{Energy Intake for $\rho_1=500$: (a) FL and (b) SMC}
\label{fig:ei2}
\end{figure}

Here, the results of $u_{1 \hspace{.5mm} min}=EI_{min}=3000$ (i.e. $\rho_1=500$) is considered and the results are given in Figs. \ref{fig:mass2}--\ref{fig:ei2}. As long as the energy intake meets the lower bound, increases in the physical activity coefficient occur (see Figs. \ref{fig:del2} and \ref{fig:ei2}). The sliding mode controller shows in Fig. \ref{fig:del2} a significantly smaller increase in $\delta$ than the feedback linearization one.

Instead of total body mass, we can consider any combination of body composition for control purposes by appropriate determination of matrix $C$ in Eq. \eqref{eq:y}. Similar results are obtained when the fat mass is considered to be controlled.

%\newpage
\vspace{1mm}
\section{CONCLUSION} \label{sec:Conclusion}
This paper has presented two model-free controllers for human weight management with physical activity as the control input. The first method is based on time-delayed feedback of the system input for approximate linearization and the second one is a switching-based controller. To estimate the daily body composition (fat, lean and extracellular fluid), a soft switching-based observer using human body weight dynamics has been proposed. This is based on daily measurements of body weight and periodic measurement of whole body composition. This paper also has addressed weight management as a control allocation problem with energy intake and physical activity coefficients as the two inputs. Based on dynamic behavior of body composition, feedback linearization and sliding mode controllers have been used to form linear algebraic equivalence of the nonlinear controllers. Then, an input-constrained nonlinear optimal controller was designed using the constrained linear least squares method.
Moreover, a subject may prefer that the start point of energy intake or expenditure is close to the current one. So this preference has been considered in the model-free set point control and optimal-model-based control. Also, it could be imposed through the reference trajectory planning of the model-free tracking control.
%------------------------------------------------------------------------------

\clearpage
\vspace{-0.75cm}
\newcommand{\BIBdecl}{\addtolength{\itemsep}{0.01mm}}   % FIXME: change this to fill the page
$\bibliographystyle{IEEEtran}$

\bibliography{obs}

\end{document}